\documentclass[reprint, aps, prl,amsmath,amssymb, twocolumn, superscriptaddress]{revtex4-1}
\usepackage[utf8]{inputenc}
\usepackage[T1]{fontenc}
\usepackage{textcomp}
\usepackage{graphicx}

\begin{document}


\title{Polarisation control of linear dipole radiation using spin-momentum locking of light}


\author{Maxime Joos}
\affiliation{Laboratoire Kastler Brossel, UPMC-Sorbonne Universités, CNRS, ENS-PSL Research University, Collège de France}
\author{Chengjie Ding}\affiliation{Laboratoire Kastler Brossel, UPMC-Sorbonne Universités, CNRS, ENS-PSL Research University, Collège de France}
\author{Vivien Loo}\affiliation{Laboratoire Kastler Brossel, UPMC-Sorbonne Universités, CNRS, ENS-PSL Research University, Collège de France}
\affiliation{ESPCI Paris, PSL Research University, CNRS, Institut Langevin}
\author{Guillaume Blanquer}\affiliation{ESPCI Paris, PSL Research University, CNRS, Institut Langevin}
\author{Elisabeth Giacobino}\affiliation{Laboratoire Kastler Brossel, UPMC-Sorbonne Universités, CNRS, ENS-PSL Research University, Collège de France}
\author{Alberto Bramati}\affiliation{Laboratoire Kastler Brossel, UPMC-Sorbonne Universités, CNRS, ENS-PSL Research University, Collège de France}
\author{Valentina Krachmalnicoff}\affiliation{ESPCI Paris, PSL Research University, CNRS, Institut Langevin}
\author{Quentin Glorieux}\affiliation{Laboratoire Kastler Brossel, UPMC-Sorbonne Universités, CNRS, ENS-PSL Research University, Collège de France}


\date{\today}

\begin{abstract}
We experimentally demonstrate that a linear dipole is not restricted to emit linearly polarised light, provided it is embedded in the appropriate nanophotonic environment.
We observe emission of various elliptical polarisations by a linear dipole including circularly polarised light, without the need for birefringent components.
We further show that the emitted state of polarisation can theoretically span the entire Poincaré sphere.
The experimental demonstration is based on elongated gold nanoparticles (nanorods) deposited on an optical nanofibre and excited by a free-space laser beam.
The light directly collected in the guided mode of the nanofibre is analysed in regard to the azimuthal position and orientation of the nanorods, observed by means of scanning electron microscopy.
This work constitutes a demonstration of the mapping between purely geometrical degrees of freedom of a light source and all polarisation states and could open the way to new methods for polarisation control of light sources at the nanoscale.
\end{abstract}

\maketitle

The wave nature of light implies that radiation phenomena are the combined manifestation of a source and its environment. Thus, properties of light sources such as power, frequency, or directivity can be tuned by confining the source in an appropriate environment. Nanophotonics aims at controlling these properties using the interaction with wavelength-scaled structures \cite{Benson} such as cavities, antennas \cite{Novotny2011} or waveguides \cite{Guoe1700007}.
These structures rely on the discretisation of the optical modes when the spatial confinement is of the order of the wavelength but until recently, they did not explicitly make use of the vectorial nature of light.
In the last decade, exploiting the strong confinement of vectorial fields has proven to be a very rich aspect of nanophotonics. Indeed, the sharp transverse confinement of vectorial fields breaks the symmetry of the intensity profile and modifies the polarisation structure of the light \cite{KIEN2004445}. When the transverse electric field varies significantly over a wavelength -- as it is the case in evanescent waves -- a longitudinal field component emerges and oscillates in phase quadrature with the transverse field. The local polarisation becomes dependent on the propagation direction of light, an effect referred to as spin-momentum or polarisation-direction locking of light \cite{VanMechelen:16} \cite{Lodahl2017}. In this context, directional emission \cite{Rodriguez-Fortuno328, Petersen67, Soellner}, non-reciprocal light-matter interaction \cite{Sayrin} and developments in nano-polarisation devices have been achieved \cite{Rodriguez-Fortuno2014, Rodriguez-Fortuno2017}.

Optical nanofibres are privileged interfaces to investigate the interaction of spin-momentum locked light with matter. In such cylindrical waveguides, the light extends outside the fibre in the form of an evanescent field and enables easy interfacing with emitters such as atoms, quantum dots \cite{Yalla2012} or plasmonic scatterers.

In this work, we exploit the strong confinement of light in optical nanofibers to demonstrate that a linear (1D) dipole can emit elliptically polarised light in the transverse plane when coupled to spin-momentum locked modes. We study how the guided light polarisation depends on the local orientation and position of the dipole and further show that the whole Poincaré sphere is accessible with this nanophotonics system.
For example, we observed emission of quasi-circularly polarised light by a linear dipole oriented at $46 \text{°}$ with respect to the propagation direction of light. We remark that this effect does not involve any birefringent component. This system constitutes a step toward a complete control of light emission at the nanoscale.

\begin{figure}[htbp]
\centering
\includegraphics[width=\linewidth]{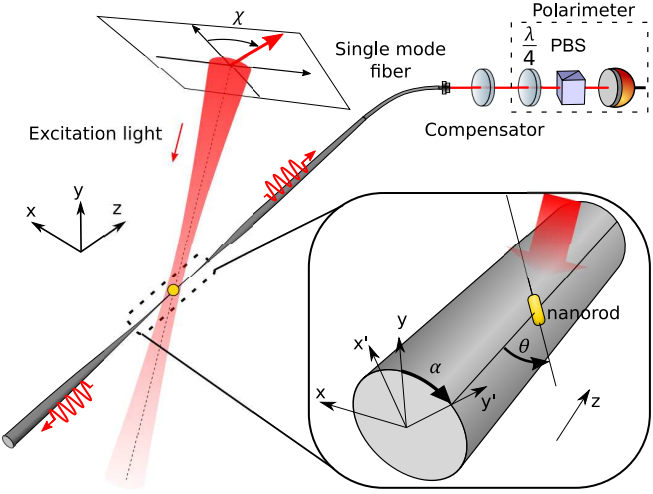}
\caption{\textbf{Experimental setup}. A single gold nanorod is deposited on the surface of an optical nanofibre. A focused laser beam contained in the $yz$-plane illuminates the particle that scatters light in the fundamental mode of the nanofibre. The nanorod lies in a plane tangent to the nanofibre surface. The normal to this plane ($y'$) makes an angle $\alpha$ with the vertical ($y$) axis. In this plane the nanorod forms an angle $\theta$ with the nanofibre ($z$) direction. The guided light is then analysed with a polarimeter allowing to measure Stokes parameters.}
\label{fig:1}
\end{figure}

Our experimental physical system is composed of a single gold nanorod deposited on the surface of an air-clad optical nanofibre as shown in figure \ref{fig:1}a. A linearly polarized, focused laser beam illuminates the nanorod particle that behaves as a linear dipole scatterer. The polarisation of the light collected in the guided mode of the nanofibre is analysed as a function of the position of the nanorod on the fibre and of its orientation, which are measured by means of a Scanning Electron Microscope (SEM).

We use optical nanofibres produced by a standard heat-and-pull process (see \cite{Hoffman}) in order to locally stretch a commercial single-mode fibre to a subwavelength-diameter of around $300\ \text{nm}$ while maintaining over $95$\% optical transmission. At our working wavelength of $637\ \text{nm}$, the nanofibre only guides the fundamental mode defined as $\text{\textbf{HE}}_{11}$ \cite{KIEN2004445}.

We now consider in more detail how an oscillating dipole couples to the fundamental mode of the nanofibre. The dipole is assumed to lie flat on the nanofibre surface at an azimuthal position denoted by the angle $\alpha$ and forms an angle $\theta$ with respect to the propagation axis $z$. We use a coordinate system ($x'y'z$) obtained by the rotation of axes $x$ and $y$ through an angle $\alpha$ (see figure \ref{fig:1}). The guided electric field $\text{\textbf{E}}_g$ emitted by the dipole can be expressed as a linear combination of the quasi-linearly polarised hybrid modes \cite{KIEN2004445, REITZ2012}
\begin{equation}
\begin{split}
\text{\textbf{HE}}_{11,x'} &= \left[\text{\textbf{e}}_{x'}(r,\varphi) + i\text{\textbf{e}}_{z}(r)\cos \varphi \right]e^{i(\beta z-\omega t)},\\
\text{\textbf{HE}}_{11,y'} &= \left[\text{\textbf{e}}_{y'}(r,\varphi) + i\text{\textbf{e}}_{z}(r)\sin \varphi \right]e^{i(\beta z-\omega t)}
\end{split}
\label{eq:0}
\end{equation}
where $\text{\textbf{e}}_{x'}(r,\varphi)$,  $\text{\textbf{e}}_{y'}(r,\varphi)$ are the real-valued transverse components and $\text{\textbf{e}}_{z}(r)\cos \varphi$,  $\text{\textbf{e}}_{z}(r)\sin \varphi$ are the longitudinal components of the guided field with main polarisation along $x'$ and $y'$ respectively, $\beta$ is the propagation constant and $\omega$ the angular frequency of the light field. This yields the guided field:
\begin{equation}
\text{\textbf{E}}_g = A\text{\textbf{ HE}}_{11,x'} + B\text{\textbf{ HE}}_{11,y'}
\label{eq:0bis}
\end{equation}
where $A$ and $B$ are complex coefficients.  $A$ and $B$ are determined by the projection of the dipole moment $\text{\textbf{d}} \propto (\sin \theta, 0, \cos \theta)$ onto the complex envelope of the hybrid modes (\ref{eq:0}) evaluated at the dipole position ($a, \pi/2$), $A = \text{\textbf{d}} \cdot \text{\textbf{e}}_{x'}(a, \pi/2)$ and $B = \text{\textbf{d}} \cdot \left[\text{\textbf{e}}_{y'}(a, \pi/2) + i\text{\textbf{e}}_{z}(a)\right]$ where $a$ is the nanofiber radius. The scalar products reduce to a single term, real for $A$ and purely imaginary for $B$ leading to the electric field:
\begin{equation}
\text{\textbf{E}}_g=  C \sin \theta\text{\textbf{ HE}}_{11,x'} + i\ D \cos \theta\text{\textbf{ HE}}_{11,y'}.
\label{eq:1}
\end{equation}
where $C$ and $D$ are real.
One first consequence of the coupling of the dipole to the nanofibre is that, whatever the orientation $\theta$ of the dipole, the guided field (\ref{eq:1}) never vanishes. When aligned along the nanofibre ($\theta = 0\text{\textdegree}$), the dipole can even radiate light in its axis direction --thanks to the nanofibre guided modes-- in strong contrast to the dipole radiation pattern in free-space. Considering now the emitted polarisation in the nanofibre: the terms of the right-hand side of (\ref{eq:1}) oscillate in phase quadrature and give rise in general, to elliptical polarisation in the transverse plane. This can be understood as follow: the dipole has non-zero overlap only with the longitudinal \textit{z}-component of $\text{\textbf{HE}}_{11,y'}$ and with the transverse \textit{x}'-component of $\text{\textbf{HE}}_{11,x'}$. The dipole excites $\text{\textbf{HE}}_{11,y'}$ through its longitudinal component and $\text{\textbf{HE}}_{11,x'}$ through its (transverse) \textit{x}'-component. For spin-momentum locked light however, transverse and longitudinal components are in phase quadrature so that the hybrid modes $\text{\textbf{HE}}_{11,x'}$ and $\text{\textbf{HE}}_{11,y'}$ will also oscillate in phase quadrature.

Let us now illustrate this mechanism with two cases: when the dipole is aligned ($\theta = 0$\textdegree) or perpendicular ($\theta = 90$\textdegree) to the nanofibre. For $\theta = 0\text{\textdegree}$, the electric field (\ref{eq:1}) of the guided light reduces to $\text{\textbf{E}}_g \propto \text{\textbf{HE}}_{11,y'}$ which corresponds to quasi-linearly polarised light along $y'$. For $\theta = 90\text{\textdegree}$, light is quasi-linearly polarised along $x'$. Between this two limiting cases, the continuous panel of polarisation ellipses whose axes are $x'$ and $y'$ is accessible. A remarkable intermediate dipole orientation, for which $|A| = |B|$ gives rise to \textit{quasi-circularly} polarised light. This can be numerically solved and leads to $\theta_{circ} = \pm 43$\textdegree for typical experimental conditions and a nanofibre diameter of $305\text{ nm}$. The dipole orientation hence defines the ellipticity of the guided light.

The azimuth $\alpha$ of the dipole on the other hand determines the major axis direction of the polarisation ellipse. This can be easily seen considering the situation $\theta = 0 \text{\textdegree}$. As already mentioned, the guided light polarisation is quasi-linear along the $y'$ direction which is, by definition, the azimuth of the dipole.

\begin{figure}[htbp]
\centering
\includegraphics[width=\linewidth]{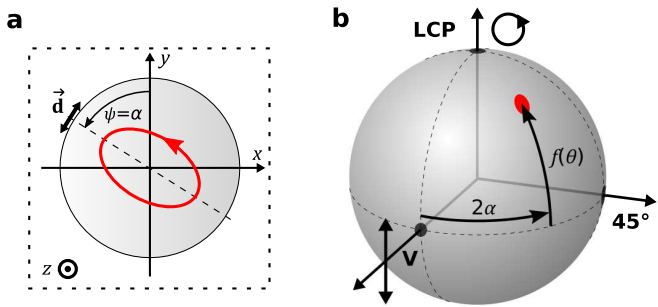}
\caption{\textbf{Mapping between dipole geometry and emitted polarisation}. \textbf{(a)} Cut of the nanofibre. The azimuth of the dipole $\alpha$ defines the orientation of the polarisation ellipse $\psi = \alpha$. \textbf{(b)} Poincaré sphere: a dipole with geometrical parameters $(\alpha, \theta)$ gives rise to a guided polarisation represented by a point with coordinate ($2\alpha, f(\theta)$) on the Poincaré sphere. V, $45\text{\textdegree}$ and LCP represent vertically linearly, linearly at $\psi = 45\text{\textdegree}$ and left-circularly polarised light respectively.}
\label{fig:5}
\end{figure}

Hence, all polarisation states are deterministically accessible through the following mapping: the nanorod azimuthal position $\alpha \in [-90\text{\textdegree}, 90\text{\textdegree}]$ determines the polarisation ellipse orientation $\psi$ or equivalently the longitude $2\alpha$ on the Poincaré sphere as shown on figure \ref{fig:5}. The nanorod orientation $\theta \in [-\theta_{circ}, \theta_{circ}]$ defines the polarisation ellipticity or equivalently the latitude on the Poincaré sphere according to a non trivial mapping $f(\theta) \approx \frac{90\text{\textdegree}}{\theta_{circ}} \theta$.

Direct measurement of the polarisation, i.e. determination of the Stokes parameters in the nanofibre region is hard to implement and we rather choose to measure the polarisation state after light exits from one end of the fibre. While light propagates in the transition region, the adiabatic tapering enables almost perfect conversion from the fundamental mode of the nanofibre to the ${LP}_{01}$ mode of the standard fibre. The polarisation however is not maintained in practice, because of the significant birefringence in fibres. In order to still map the hybrid mode basis onto an accessible paraxial mode basis, we use a uniaxial birefringent plate mounted in the form of a Bereck compensator to compensate the birefringence effect of the fibre.

In our experiment, we use single gold nanorods that, under suitable illumination, behave as dipole scatterers whose dipole orientations are along the rods longitudinal axes. The deposition of a single nanorod on the nanofibre surface is performed by touching it with a small droplet of a commercial colloidal particle dispersion (Nanocomposix). The nanorods were chosen so that their longitudinal surface plasmon resonance (LSPR) matches our laser wavelength of $637\text{ nm}$ when deposited on the nanofibre. We hence work with nanorods of average aspect ratio of $2.6$ yielding a LSPR centred around our working wavelength and with a size of $17 \pm 1\ \text{nm}$  in diameter and $ 45 \pm 6\ \text{nm}$ in length, considered small compared to the nanofibre geometry.

We model a nanorod as an anisotropic scatterer on which an exciting electric field $\text{\textbf{E}}_{exc}=(E_L, E_T)$ induces a dipole moment $ \text{\textbf{p}} = (\alpha_L E_L, \alpha_T E_T)$ where $\alpha_L$ and $\alpha_T$ are the longitudinal and transverse polarisabilities of the rod, respectively. Close to the LSPR,  $\text{Im}(\alpha_L) \gg \text{Im}(\alpha_T)$ and the transverse component of the dipole moment is strongly suppressed compared to the longitudinal component. We can hence suppose that, as long as the excitation field does not oscillate too perpendicularly with respect to the rod, in which case $\alpha_L E_L$ might be comparable to $\alpha_T E_T$, the dipole is induced along the nanorod.

\begin{figure}[htbp]
\centering
\includegraphics[width=0.9\linewidth]{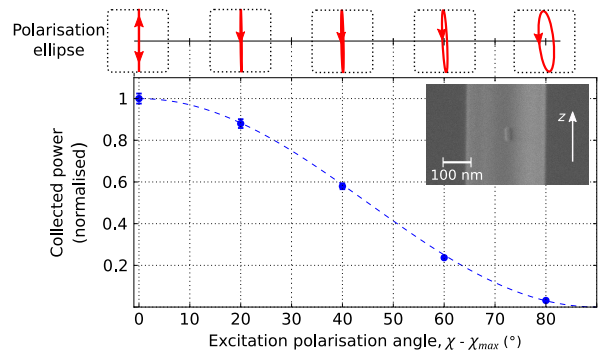}
\caption{\textbf{Nanorod optical response to rotating linear polarisation excitation}. The lower graph represents the normalised collected power in the nanofibre as the excitation polarisation angle $\chi$ rotates over $80\text{\textdegree}$, starting at $\chi = \chi_{max}$ for which the scattered signal is maximum. The measured collected power is fitted to a Malus law. \textbf{Inset:} SEM image of the measured nanorod on the nanofiber. The upper plots represent the measured polarisation ellipses for different excitation polarisation angles. The guided polarisation is almost independent of $\chi$ which constitutes a consistency check of our assumption that the dipole tends to form along the rod.}
\label{fig:2}
\end{figure}

To verify this crucial assumption experimentally, we rotated the excitation polarisation while monitoring the power scattered by the nanorod into the nanofibre. We observed a sinusoidal dependence corresponding to a Malus law as shown in figure \ref{fig:2}. The beam polarisation angle for which the scattered signal is maximum is referred to as $\chi_{max}$. In contrast to the collected power into the nanofibre, the guided polarisation is almost unchanged when turning the excitation polarisation. This validates the assumption that the dipole tends to align along the rod, independently of the excitation polarisation. In the following experiments, we will always maximise the scattered signal by adjusting the beam polarisation orientation to the value $\chi_{max}$ prior to acquiring the guided polarisation state. The intrinsic diffusion of the excitation beam by the bare nanofibre is lower than $0.5\ \%$ of the overall scattered signal and can hence be neglected.

The experimental procedure consists in: (1) depositing a nanorod, (2) compensating the fiber birefringence, (3) aligning the excitation polarisation along the rod and finally (4) recording the guided polarisation state. It is repeated several times at different locations along the nanofibre. A special care is dedicated to recording the exact relative positions of the particles with respect to each other. This is meant to facilitate the finding of the nanorods when we ultimately observe the nanofibre with a SEM. Then, we discriminate the various depositions, eliminate the clusters and rods with odd shape to record only the azimuth $\alpha$ and the orientation $\theta$ of \textit{proper} single nanorods. We note that experimentally, we never observed rods with orientation $|\theta| \gtrsim 60\text{\textdegree}$ which we understand as the tendency for the nanorod to maximise its contact surface with the nanofiber.

\begin{figure}[htbp]
\centering
\includegraphics[width=\linewidth]{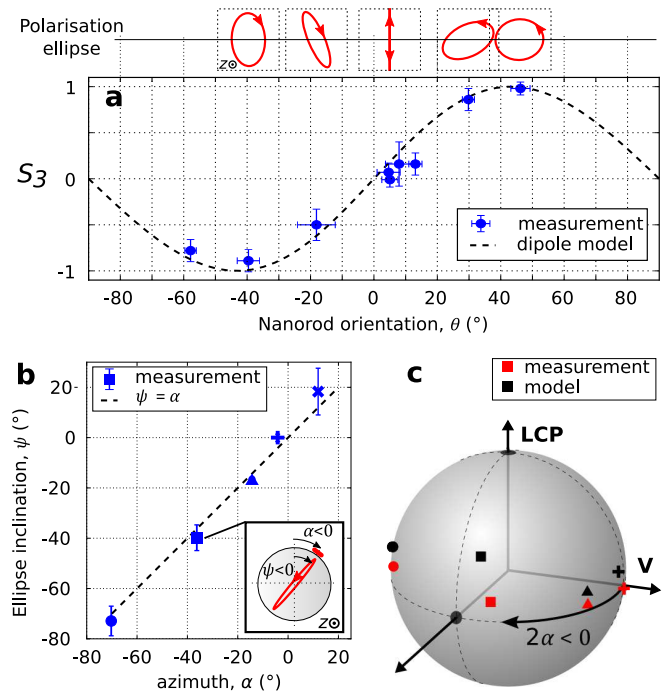}
\caption{\textbf{Measured state of polarisation as a function of the orientation $\theta$ and azimuth $\alpha$ of the nanorod}. \textbf{(a)} Ellipticity as a function of $\theta$. Graph : measured Stokes parameter $S_3$ for 9 different nanorods compared to the expected dipole model. We plot the measured polarisation ellipse for characteristic nanorods. \textbf{(b)} Measured ellipse orientation $\psi$ for 5 nanorods with different azimuth. The inset represents a cut of the nanofibre showing the azimuth of a specific nanorod and the associated polarisation ellipse. \textbf{(c)} Poincaré sphere representation of the measured polarisation for 4 nanorods from \textbf{(b)}.}
\label{fig:4}
\end{figure}

In our model of a linear dipole coupled to nanofiber modes, we predicted that the dipole orientation determines the polarisation ellipticity. In order to express this dependence and compare our measurement to the model, we consider the normalised Stokes parameter $S_3$ as a function of the dipole/nanorod orientation $\theta$ as shown in figure \ref{fig:4}a. $S_3$ ranges from $-1$ to $1$ and provides us immediate information on the \textit{degree of circular polarisation} -- defined as $|S_3|$ -- and on the sense of circulation of the electric field in the transverse plane, related to the sign of $S_3$. The figure also shows the measured corresponding polarisation ellipses for five characteristic nanorods, whose orientation vary from $-40\text{\textdegree}$ to $+46\text{\textdegree}$. As input parameters for the model, we used a the nanofibre diameter extracted from SEM images (2a = $305 \pm 7\text{ nm}$), the working wavelength of $637\text{ nm}$, $1.457$ and $1.000$ for the refractive index of silica and air respectively and we take $9\text{ nm}$, which corresponds to the radius of a nanorod for the distance of the dipole to the nanofibre surface.

Our measurements of $S_3$ fit well our dipole model and present the characteristic behaviour presented above namely that, in the range $|\theta| \lesssim \theta_{circ}$, the degree of circular polarisation grows with $|\theta|$ as the two components in (\ref{eq:1}) reach comparable amplitude. The case of circular polarisation is expected at $\theta_{circ} = \pm 43\text{\textdegree}$ and experimentally, we observed almost purely circularly polarised light for a nanorod oriented at $\theta = 46 \pm 3 \text{\textdegree}$. In contrast, nanorods nearly aligned along the nanofiber (small $\theta$) induce a low degree of circular polarisation and hence a narrow polarisation ellipse.
We also notice that the measured handedness of the polarisation fits to our model as positive (negative) $\theta$ gives rise to counter-clockwise (clockwise) circulation of the electric field in the transverse plane.

We now consider the relation between the azimuthal position $\alpha$ of the nanorod and the polarisation ellipse orientation $\psi$. This is easily observable for rods whose polarisation ellipse is narrow, i.e. whose orientation $\theta$ forms a small angle with the nanofibre. We hence consider five rods with orientations $\theta$ lying between $-20\text{\textdegree}$ and $20\text{\textdegree}$. Figure \ref{fig:4}b shows the measured ellipse orientations as a function of the nanorod azimuth. The measured orientations of the polarisation ellipses tend to follow the azimuthal positions of the nanorods.

Thus our experiments show that the measured polarisation states as a function of the orientation and position of the nanorod are in good agreement with our linear dipole model. This constitutes an experimental demonstration that this system can generate all possible polarisation states.

In conclusion, we used gold nanorods deposited on a nanofibre to demonstrate that a linear dipole can radiate elliptically polarised light when coupled to spin-momentum locked modes. We showed that this system can, in principle, generate all possible polarisation states. The azimuth position of the rod controls the inclination of the polarisation ellipse while its orientation with respect to the nanofibre defines the ellipticity. This constitutes a demonstration of the mapping between purely geometrical degrees of freedom of a light source and polarisation states. This system opens a new way for controlling the polarisation of light sources at the nanoscale without involving birefringent components or magnetic fields.

The authors would like to thank Clément Sayrin for his helpful suggestions throughout this work, Imène Estève for her decisive support on the SEM platform and Sébastien Bidault for fruitful discussions.

\section*{Funding} This work is supported by the Emergence program from Ville de Paris and by PSL Research
University in the framework of the project COSINE.

\bibliography{bibliography}


\end{document}